# Impact of Ge substrate Thicknesses and Epitaxy Growth Conditions on the Optical and Material Properties of Ge- and GaAs-based VCSELs

Wenhan Dong[1,2], Zeyu Wan[1], Yun-Cheng Yang[3], Chao-Hsin Wu[3], Yiwen Zhang[2], Rui-Tao Wen[2*], Guangrui Xia[1*]

[1] Department of Materials Engineering, University of British Columbia, Vancouver, Canada
[2] Department of Materials Science and Engineering, Southern University of Science and Technology (SUSTech), Shenzhen, China
[3] Graduate School of Advanced Technology, National Taiwan University, Taipei 106319, Taiwan
[*] Corresponding authors: wenrt@sustech.edu.cn, gxia@mail.ubc.ca

**Abstract:** We present a comparative study of the optical and material property dependences of VCSELs on Ge or GaAs substrate thicknesses and epitaxy process conditions. It was found that adjusting the Ge substrate thickness and optimizing the epitaxy process can shift the stopband center and cavity resonance wavelength by several nanometers. Ge-based VCSELs exhibit improved epitaxial uniformity, smaller deviations from design specifications, reduced stoichiometry variations, and strain magnitudes comparable to those of GaAs-based counterparts. In the selected 46.92 μm² sample area, no defects were observed in the quantum well (QW) regions of Ge-based VCSELs, and the threading dislocation density (TDD) was measured to be below $2.13 \times 10^6$ cm$^{-2}$. These results highlight the potential of Ge substrates as promising candidates for advanced VCSELs.

## 1. Introduction

Vertical cavity surface-emitting lasers (VCSELs) have many advantages such as high power efficiency, low beam divergence, narrow spectrum, higher spectral and temperature stability, low-cost testing and packaging, ease in testing, and scalability. They are widely used in modern optoelectronic applications such as short-reach optical communication, sensing, and imaging. The growing demands from optical networking, consumer electronics, and Light Detection and Ranging (LiDAR) technology are driving rapid expansion of the VCSEL industry. VCSEL manufacturing is scaling up globally, requiring continuous improvements in manufacturing processes to achieve higher yields and quality [1, 2]. The VCSEL market is expected to double from $1.2 billion in 2021 to $2.4 billion by 2026, with a compound annual growth rate of 13.6% [3, 4].

At present, VCSELs are primarily fabricated on 4 and 6" GaAs wafers [5]. Although AlGaAs-based structures on GaAs are well established, challenges remain in scalability, yield, and cost. Strain-induced wafer bowing limits the use of larger GaAs wafers, while their fragility constrains expansion beyond 8". VCSEL epitaxial structures typically consists of upper and lower distributed Bragg reflectors (DBRs) made of alternating $Al_xGa_{1-x}As$ and $Al_yGa_{1-y}As$ DBR layers ($x < y$), which constitute most of the total thickness. Germanium (Ge) offers a promising alternative, with a lattice constant (5.658 Å) between those of GaAs and AlAs, reducing lattice mismatch and associated strain [6]. This enables growth on larger, more stable 12" wafers. Ge also provides greater mechanical strength, lower threading dislocation densities, and improved post-growth stability, supporting enhanced scalability and cost efficiency in VCSEL production [7].

In 2021 and 2023, IQE plc announced the successful epitaxy of 6 and 8" Ge-based VCSEL wafers [8, 9]. However, specific details on epitaxy design, growth processes, conditions, and layer structure remain undisclosed. In 2021, researchers from the University of British Columbia, National Taiwan University, and Fraunhofer Institute of Solar Energy fabricated bulk Ge-based n-doped distributed Bragg reflectors (n-DBRs) and half-VCSELs, followed by successful full VCSEL epitaxy in 2022 [10]. Ge-substrate VCSELs developed independently by the same team demonstrated successful lasing in 2023, achieving 3dB bandwidths of 16.4 GHz at room temperature and 16.1 GHz at 85°C [11]. These bulk Ge-based AlGaAs VCSELs have been shown much less wafer bow and warp, better surface roughness, and comparable electrical and optical performance compared with the mainstream GaAs counterparts [12]. These advantages highlight the potential of Ge-based VCSELs, supporting enhanced scalability, efficiency, and high-volume manufacturing.

Bulk Ge wafers have traditionally been used in solar cell applications, and there are currently no established standards for Ge substrate thickness for laser applications. As Ge-based VCSELs remain in the early stages of development, the impact of Ge substrate parameters such as thickness, thermal transfer, initial wafer bow or warp, and process optimization, on epitaxial growth has not yet been systematically explored. Meanwhile, high-resolution transmission electron microscopy (TEM) offers powerful nanoscale imaging capabilities suited for analyzing the quantum well (QW) region of VCSELs. Despite its advantages, TEM has not been previously used for detailed structural assessment of VCSEL QWs, and comparative nm-scale studies of Ge- and GaAs-based VCSELs are absent from the literature. This study addresses both of these important gaps by combining structural and optical characterization techniques to evaluate the impact of Ge substrates and to conduct direct comparisons with conventional GaAs-based VCSELs.

## 2. Sample preparation and experiment method

The full VCSEL epitaxy, designed for operation at 940 nm, were grown using a production-scale metal-organic chemical vapor deposition (MOCVD) at LandMark Optoelectronics, with detailed structural parameters described in Ref. [12]. In this study, the bulk substrates were 4" Ge and GaAs wafers of varying thicknesses: 330 μm and 425 μm for Ge, and 450 μm and 625 μm for GaAs, as summarized in Table 1. The Ge wafers fabricated by Umicore are double-sided polished with a 50-100 nm SiN film to prevent Ge sublimation. These wafers were separated as two groups, where full VCSEL epitaxy with two different process optimizations were conducted. Process 1 was optimized for sample Ge330 (330 indicates the Ge thickness is 330 μm), while Process 2 was optimized for sample GaAs625. These two wafer thicknesses were chosen because the thinnest Ge wafers with good VCSEL performance were 330 μm Ge wafers, and 625 μm has been a common 4-6" GaAs wafer thickness [13]. We expected that Process 1 and 2 should produce good-performing Ge-based and GaAs-based VCSELs respectively so that the subsequent comparisons were fair.

The epitaxy process optimizations were conducted using surface temperature (measured by thermoreflectance) and the precursor mass flow rates as key tuning parameters. These parameters were carefully tuned to produce smooth surface morphology, effective dopant incorporation, uniformity, and good alignment of the stopband characteristics and QW emission wavelength with the design values.

|  | Process Optimization | Samples Grown |
|---|---|---|
| Process 1 | Optimized for Ge330 | Ge330, Ge425, GaAs450, GaAs625 |
| Process 2 | Optimized for GaAs625 | GaAs450, GaAs625 |

Table 1. Two epitaxy processes and the corresponding samples grown using the processes. The numbers in the sample names indicate the substrate thickness in microns.

Bow and warp data for Ge- and GaAs-based VCSEL wafers were obtained using the FLX-2320-S system from Toho Technology. The system measures surface curvature by directing a laser at known incident angles and detecting the reflected beam with a position-sensitive photodiode. By scanning across the wafer, curvature data are collected at multiple points to quantify bow and warp. In this work, the top DBR stopbands were characterized using a Nanometrics RPM Blue wafer laser measurement tool, calibrated with a standard silver mirror before measurement. The lasing and the L-I-V data comparison after device fabrication were published in [12], and will not be discussed here. The material properties were studied by TEM analysis. Cross-sectional VCSEL samples were prepared using the FEI Helios NanoLab™ 600i, equipped with a high-resolution SEM column and a focused ion beam (FIB) with a gallium ion source, enabling precise site-specific nano-milling. A protective platinum coating was applied during FIB processing to prevent surface damage. High-resolution TEM imaging was performed using a Thermo Scientific Talos™ F200X G2 TEM to analyze the quantum well structure and material properties of the VCSELs.

Layer thickness was determined from high resolution TEM images by analyzing pixel intensity profiles along image rows. Layer boundaries were identified at local maxima of the intensity slope, and thickness was calculated from the pixel distance between these boundaries. TEM Energy Dispersive X-ray (EDX) spectroscopy was used to analyze elemental composition and distribution in the QW regions. Atomic fraction of detected elements in the QW layers were recorded for compositional analysis.

## 3. Results and discussion

### 3.1 Effect of the substrates on the optimal growth temperatures

To monitor the growth surface temperature during epitaxy, thermoreflectance was employed. This technique detects small changes in surface reflectance of layers caused by temperature-induced variations in optical constants, allowing for real-time, non-contact monitoring of the wafer's surface temperature with high accuracy throughout the growth process. Regardless of the substrate type, high-quality epitaxy was consistently achieved with very close surface temperatures on both GaAs and Ge substrates with small tuning in gas flows. This confirms that surface temperature is a primary determinant of epitaxial quality. Notably, the optimized setting temperatures for Ge wafers to reach such surface temperatures were 15–20 °C lower than those for GaAs, and variations in Ge substrate thickness had a minimal impact on the optimized setting temperatures.

This difference can be explained by the thermal contact and the heat transfer to the wafers. Firstly, Ge wafers were double-side polished, allowing more efficient thermal contact with the susceptor pocket and enhancing heat conduction to the wafer surface. In contrast, the GaAs wafer backsides were not polished, resulting in less effective thermal contact and thus requiring a higher heating power to achieve the same surface temperature. Secondly, Ge wafers exhibit better wafer flatness. As shown in Figure 1, the wafer map of the Ge330 sample (Fig. 1(a)) appears more uniform with a less peak-to-valley (PV) value, while the GaAs625 sample (Fig. 1(b)) shows a clear dome-shape caused by more lattice mismatch.

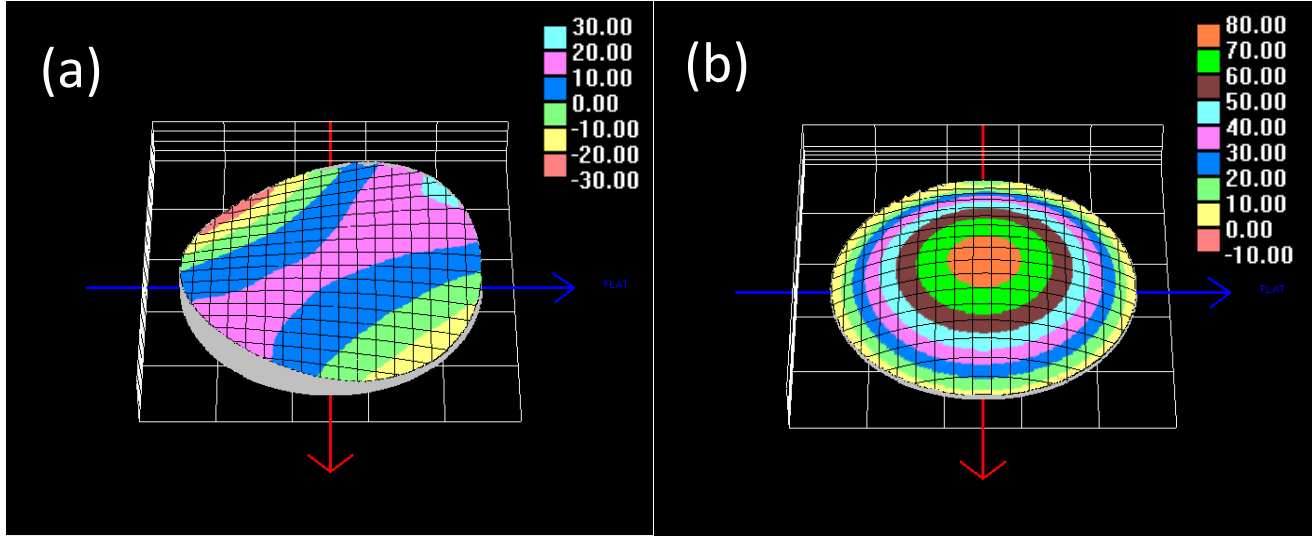

Figure 1. Wafer bow/warp maps of (a) Ge330 wafer grown by Process 1, and (b) GaAs625 wafer grown by Process 2. The side bar values are in microns.

As shown in Table 2, both Ge wafers exhibited relatively smaller average wafer warp (3.77 to 10.28 µm), and smaller PV warp values ranging from 50 to 63 µm, while four GaAs wafers showed significantly larger average wafer warp (36.59 to 63.89 µm) and PV warp values from 78 to 135 µm. Larger PV warp reflects more pronounced deformation, which may reduce physical conformity with the susceptor surface and degrade local thermal contact. Therefore, both backside polishing and reduced PV warp contribute to a lower thermal contact resistance at the wafer–susceptor interface, enabling more efficient heat transfer during epitaxy in Ge wafers.

| Wafer/ | Average wafer warp (µm) | PV wafer warp (µm) |
| --- | --- | --- |
| Ge330 (Process 1) | 3.77 | 50.30 |
| Ge425 (Process 1) | 10.28 | 62.63 |
| GaAs450 (Process 1) | 62.37 | 132.69 |
| GaAs625 (Process 1) | 37.23 | 78.31 |
| GaAs450 (Process 2) | 63.89 | 135.41 |
| GaAs625 (Process 2) | 36.59 | 76.57 |

Table 2. The wafer warp characteristics of the four wafers.

### 3.2 Thickness effect on the stopbands of the top DBRs

The reflectance spectrum is an effective approach for assessing the design and epitaxy quality of VCSEL structures. It measures the intensity of reflected light across different wavelengths.

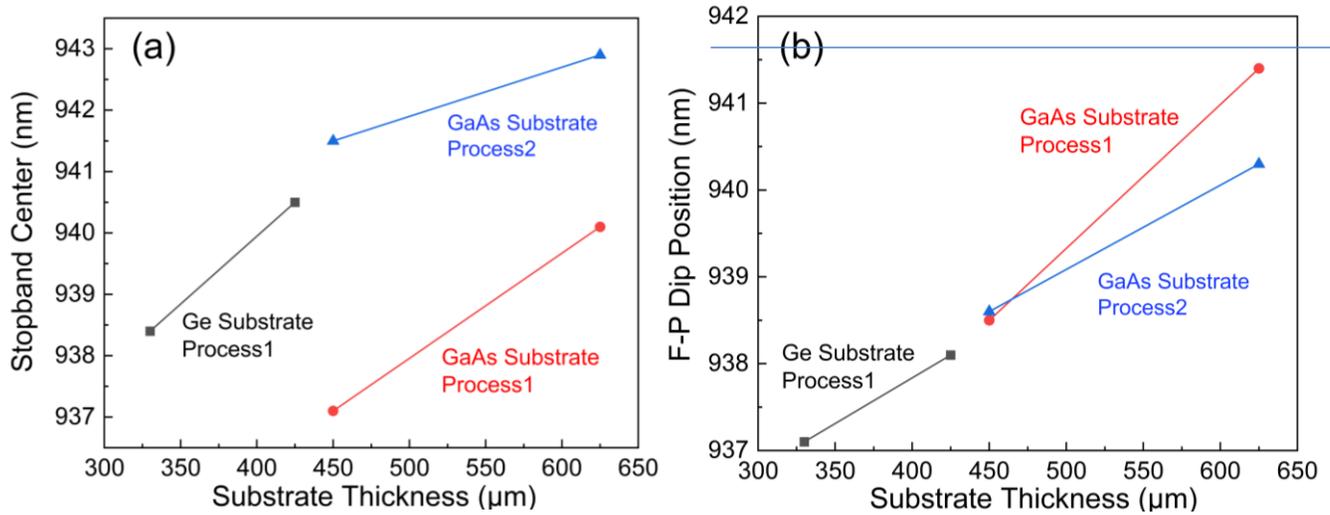

Figure 2. The substrate thickness and process optimization impacts on (a) the stopband center and (b) the Fabry-Perot dip wavelengths.

The stopband center and Fabry–Pérot dip wavelengths were extracted from room-temperature reflectance spectra measured at the wafer center. As shown in Figure 2, both Ge and GaAs wafers exhibit stopband centers closely aligned with the 940 nm design specification, indicating high-quality epitaxial growth. A measurable red shift was observed in both spectral features with increasing substrate thickness. For Ge wafers (Process 1), increasing the substrate thickness from 330 μm to 425 μm (a 28.8% increase) resulted in a center wavelength shift from 938.4 nm to 940.5 nm (0.22%) and a Fabry–Pérot dip shift from 937.1 nm to 938.1 nm (0.11%). For GaAs wafers (Process 2), increasing the thickness from 450 μm to 625 μm (38.9%) led to a center wavelength shift from 941.5 nm to 942.9 nm (0.15%) and a dip shift from 938.6 nm to 940.3 nm (0.18%).

Initially, wafer bow and warp as well as thermal conductivity differences were considered as potential causes of this thickness-dependent shift. The thermal conductivities of Ge (0.58 W/cm·°C) and GaAs (0.55 W/cm·°C) differ by only approximately 5.5%. As discussed in 3.1, optimal surface growth temperatures for all substrates are similar. This thermal conductivity difference only impacts how easy the optimal surface growth temperatures can be achieved, and should not be relevant.

Wafer bow and warp were also examined for their potential impact. As shown in Table 2, Ge wafers exhibited a 172% increase in average warp with increasing thickness (from 3.77 μm for Ge330 μm to 10.28 μm for Ge425 μm), while GaAs wafers showed a 41% decrease (from 63.89 μm for GaAs450 to 37.23 μm for GaAs625). Despite these opposite trends, both Ge and GaAs wafers exhibited a similar red shift with increasing thickness. Furthermore, both Ge wafers exhibited low average warp values, approximately 10 μm or less, indicating minimal influence on epitaxial quality. Meanwhile, GaAs wafers exhibited much higher average warp values, which were over 60 μm in thinner samples, yet showed a comparable degree of red shift. The presence of red shift in both low-warp Ge and high-warp GaAs wafers, as well as the fact that Ge and GaAs exhibited opposite trends in average warp with increasing thickness, collectively lead to the conclusion that average wafer bow and warp are not the dominant factors governing the spectral shift.

Consequently, these factors alone do not fully explain the observed thickness-dependent spectral shifts, indicating that additional mechanisms may contribute. After thorough analysis of reactor design and growth kinetics, we conclude that the spectral shift mainly results from enhanced gas flow and improved thermal uniformity during epitaxy. With the susceptor pocket depth fixed at 575 μm, thicker substrates make more effective contact within the gas ambient in the reactor, improving the interaction with gas flow and heat distribution. This results in more uniform layer deposition and a slight increase in overall layer thickness, thereby increasing the optical path length and causing the observed red shift. These findings highlight the importance of the substrate thickness, backside polishing, wafer pocket design, and the epitaxial conditions, offering insights into precise spectral control and process optimization for scalable Ge-based VCSEL fabrication.

### 3.3 Investigation of the VCSEL Structures by TEM Images

To investigate substrate effects on VCSEL QW epitaxy growth and analyze material properties, cross-sectional TEM samples of the full VCSEL structure were prepared and examined by using high-resolution TEM (HRTEM) bright-field imaging. As shown in Fig. 3, the bright-field image of the Ge-based VCSEL sample (Ge330) clearly reveals the sequential structures of the top Distributed Bragg Reflector (DBR), the quantum well, and the bottom DBR.

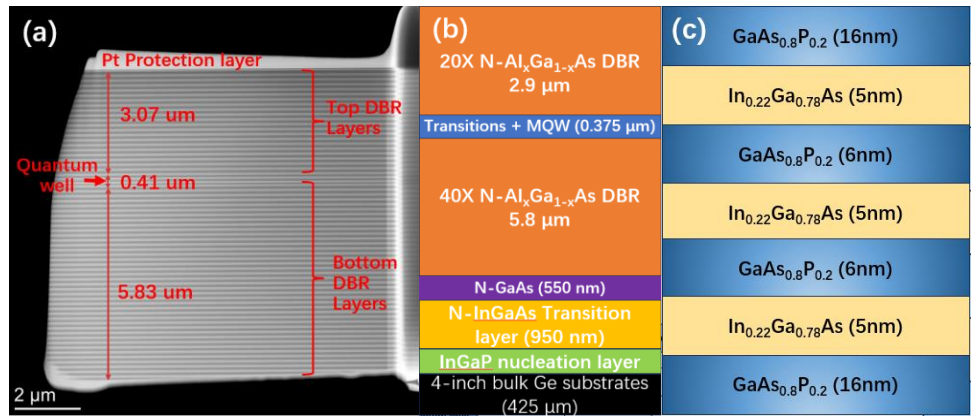

Figure 3. Comparison of (a) the TEM bright field image of Ge-based VCSEL cross-section, (b) the design schematic of Ge-based VCSEL (the thickness of Ge substrate is not to scale) and (c) the design schematic of quantum well structure.

The cross-sectional analysis indicates that the thickness deviations from the designed values (Fig. 3(b)) are within an acceptable range for epitaxial growth, with the top DBR deviating by 5.86%, the quantum well by 9.3%, and the bottom DBR by 0.52%. These results demonstrate good consistency between the fabricated structure and the intended design.

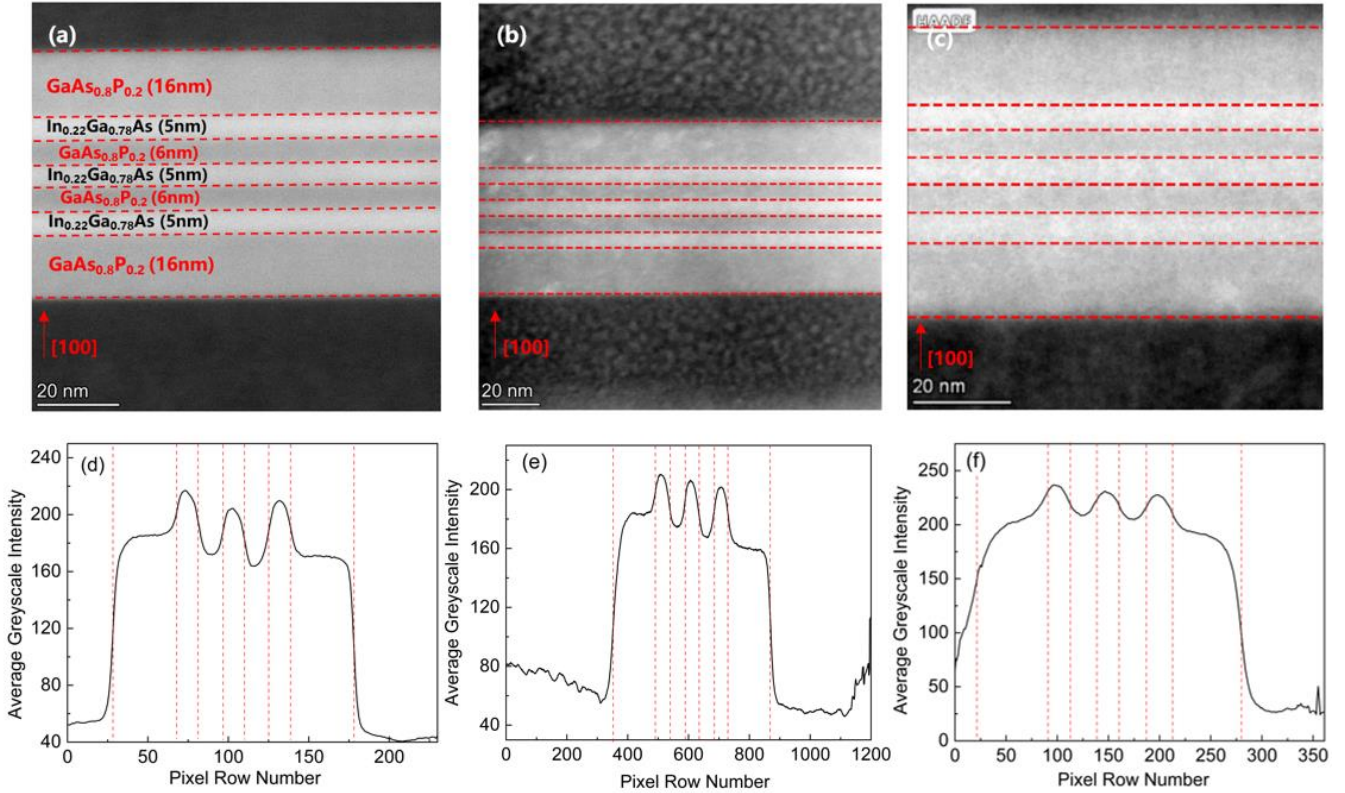

Figure 4. High-Resolution TEM bright field image of quantum well region in (a) Ge425 T4B1, (b) Ge330 T4A4 and (c) GaAs625, and the average greyscale intensity of the TEM image of (d) Ge425 T4B1, (e) Ge330 T4A4 and (f) GaAs625 along rows with boundaries labeled as red lines

To extract thickness information across a broader sample set, grayscale pixel intensity profiles were plotted from the TEM images, as shown in Fig. 4. Abrupt intensity transitions, which correspond to layer boundaries, were detected by identifying local extrema in the slope of the intensity curve. Red dashed lines mark these transitions, and the pixel distances between them were used to determine individual layer thicknesses.

The measured thicknesses were then compared with the intended design values in Fig. 3(c), and the deviations for all samples are summarized in the accompanying table. Most Ge-based samples exhibit smaller deviations than their GaAs-based counterparts. In particular, Ge330 shows minimal mean, minimum, and maximum deviations from design specifications, indicating more consistent epitaxial growth. This improved thickness control in Ge-based samples can be attributed to the enhanced thermal contact and gas flow dynamics discussed in Sections 3.1 and 3.2. The double-side polished Ge wafers provide more efficient heat transfer during growth, while their lower peak-to-valley warp ensures better contact with the susceptor. These advantages promote a more stable and uniform growth environment, resulting in reduced thickness variation and reinforcing the suitability of Ge substrates for precise epitaxial fabrication.

| Deviation | GaAs625 | Ge425 | Ge330 |
| --- | --- | --- | --- |
| Max Deviation (%) | 10.4 | 7.70 | 5.03 |
| Min Deviation (%) | 0.33 | 0.45 | 0.11 |
| Mean Deviation (%) | 4.35 | 4.10 | 3.34 |

Table. 3 Deviations of the Ge-based and GaAs-based QW layer thicknesses from the target thicknesses.

## 3.4 EDX analysis of quantum well regions in VCSEL

Elemental distribution in the QW region directly affects band alignment and carrier confinement, leading to shifts in the center wavelength of VCSELs. Deviations from stoichiometry alter strain and composition, impacting optical gain and emission stability. To identify the elemental distribution within the quantum well regions, EDX spectroscopy was performed, and the corresponding EDX mapping for the Ge330 VCSEL sample was obtained. As illustrated in Fig. 5, the elements In and P, along with Al and Ga, show complementary distributions in the quantum well region. Discrete layers are clearly visible in the EDX signal mapping of the In and P elements, which align well with the design layout presented in Fig. 3(c).

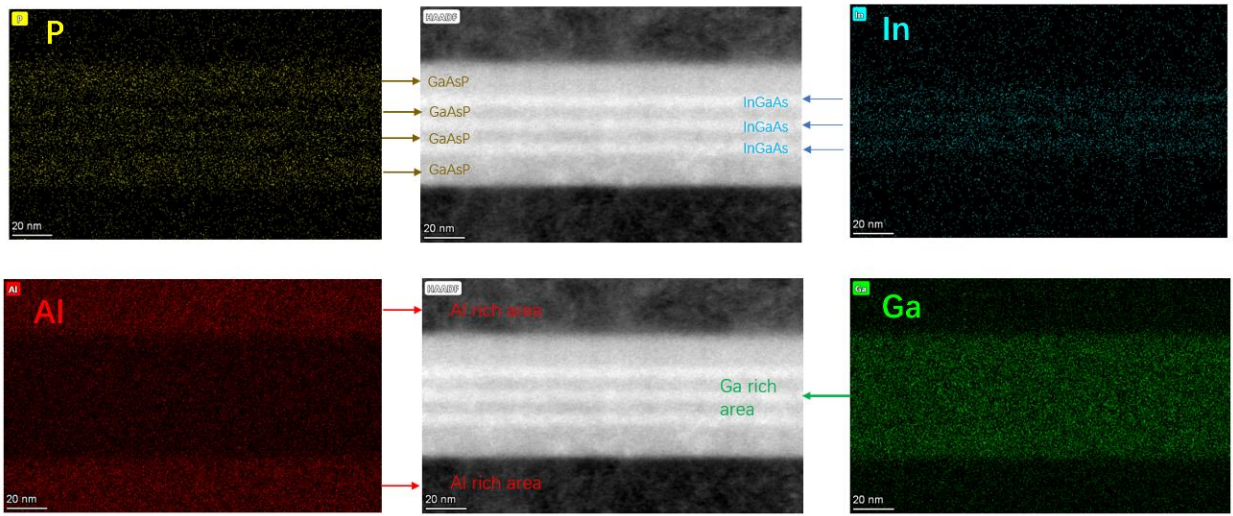

Figure 5. EDX mapping of the quantum well regions in Ge-VCSEL with different element signals

The elemental composition was analyzed through layer-by-layer EDX scans within the QW region. Due to the effects of Ga ion beam bombardment from FIB, the atomic fraction percentages of In and Ga were not considered. The QW region includes epitaxially grown $GaAs_{0.8}P_{0.2}$ in layers 1, 3, 5, and 7 (Fig. 3 (c)), where phosphorus (P) and arsenic (As) occupy the same lattice sites. The calculated P-to-total P and As ratio for each individual layer, as well as the average ratio for different samples, is shown in Fig. 6 below, with the stoichiometry marked by a purple dot-dash line at 20%. The Ge330 sample exhibits an average ratio of 19.21% shown as black dash line, with the deviation from stoichiometry of 0.79%. The average ratio of the GaAs625 is 21.28% as blue dash line, and the deviation from stoichiometry is 1.28%. The Ge425 as red dash line has an average ratio of 18.87% with an intermediate deviation of 1.13%. Overall, the EDX elemental analysis shows that the Ge samples exhibit closer alignment with ideal stoichiometry and, in most cases, smaller deviations compared to the GaAs samples. The slight deviations in elemental composition for both Ge and GaAs samples can lead to the red shifts observed in the stopband center and Fabry-Perot dip wavelengths from 940 nm (Fig. 2), suggesting a potential influence of composition variations on optical properties. However, due to the small sample count, more samples and statistical analysis are needed to confirm this finding.

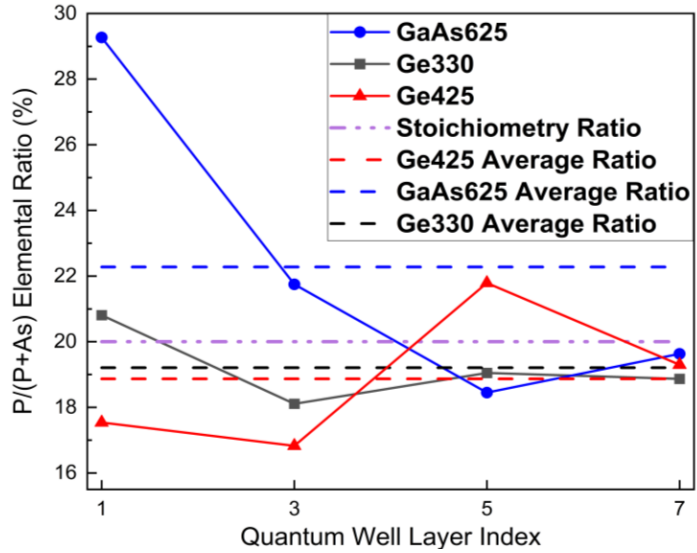

Figure 6. P/(P+As) elemental ratios across QW layers and the average ratios of Ge and GaAs substrate samples.

**3.5 Strain and defect density of quantum well regions in VCSEL**

Strain distribution and defect density in the QW region of VCSELs influence optical gain, emission wavelength, and reliability. Strain alters band alignment, while defects cause nonradiative recombination, reducing efficiency and lifespan. To investigate the strain and stress distribution in quantum well regions and study the band gap modification based on strain changes, high-resolution TEM images were analyzed using Geometrical Phase Analysis (GPA) method in Gatan Digital Micrograph. GPA measures atomic displacements by analyzing phase shifts in high-resolution TEM images, enabling the generation of atomic-scale strain maps [14]. To perform strain analysis on the selected TEM area, two spots on its Fast Fourier Transform (FFT) pattern will be chosen, corresponding to the reference planar directions. The planar spacing of these crystal planes will set as the reference for strain calculation. By comparing the actual atomic distances to the reference spacing, lattice distortion in different regions can be determined, allowing for strain mapping.

Figure 7 presents the strain mapping of the quantum well region for (a) Ge330 (Ge-based) and (b) GaAs625 (GaAs-based) samples. The red regions indicate tensile strain in the InGaAs layers, while the green regions represent compressive strain in the GaAsP layers. Strain affects the band gap energy, with tensile strain in InGaAs reducing the band gap and improving carrier confinement, while compressive strain in GaAsP increases the band gap and enhances the barrier between quantum well layers [15]. For Ge330 The strain magnitudes in the Ge-based samples are comparable to those in the GaAs-based samples, with detected strain approximately 1% for both substrates.

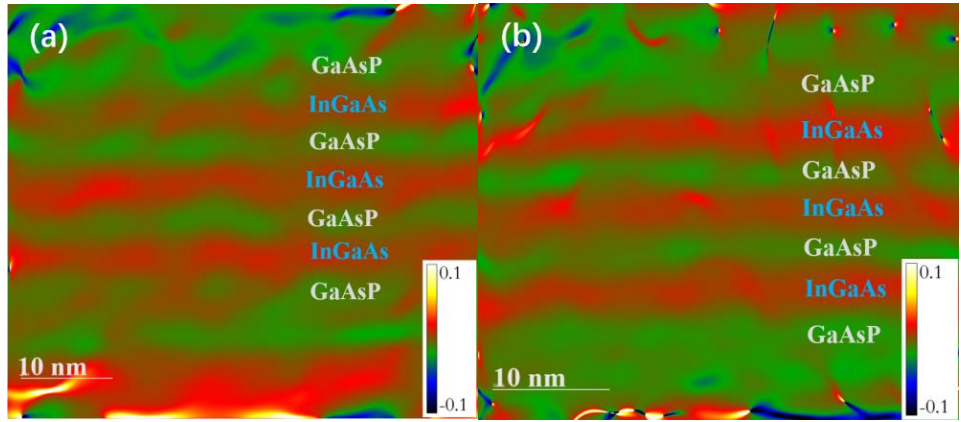

Figure 7. Strain Mapping of the Quantum Well Region in (a) Ge330 (Ge-based) and (b) GaAs625 (GaAs-based) Samples

To investigate the defect density in the QW regions of the Ge substrate sample, plan-view TEM (PVTEM) samples were prepared for defect observation. Fig. 8 shows the bright field and dark field PVTEM images of the QW region in the Ge substrate sample, with an area of approximately 36 μm². Additional PVTEM samples were prepared, covering a total sample area of 46.92 μm² for analysis. All PVTEM samples reveal that the Ge-based quantum wells exhibit no observable defects, with dislocation densities calculated to be less than $2.13 \times 10^6$ cm$^{-2}$, comparable to those in GaAs-based epitaxial layers.

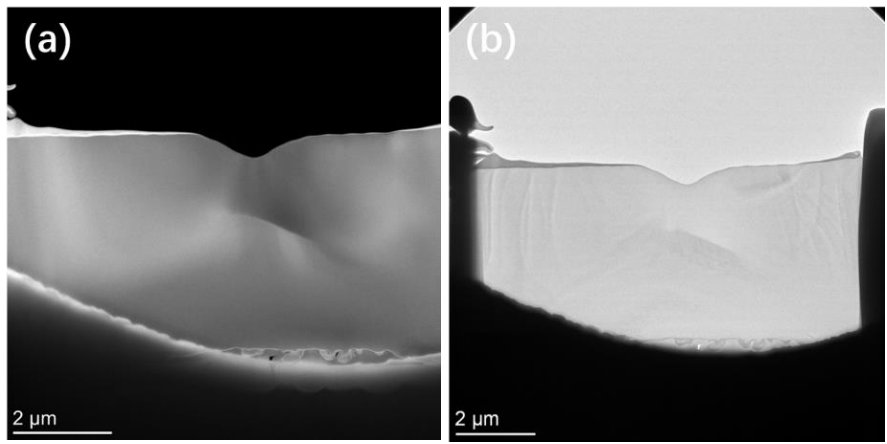

Figure 8. (a) the dark field and (b) bright field PVTEM images of the QW region in the Ge330 sample

## 4. Conclusions

This study investigates VCSELs on Ge and GaAs substrates, focusing on the impact of the substrate thickness, substrate type, and process optimization on the epitaxy quality and the optical performance. The surface temperatures for optimal epitaxy quality were similar for all wafers, while the corresponding setting temperatures for Ge wafers to reach such surface temperatures were 15–20 °C lower than those for GaAs wafers. This can be explained by the much less warped Ge wafers with backside polishing, which provided better thermal contacts to the susceptors in the epitaxy grow tool. The top DBR stopband center and Fabry-Pérot dip of both Ge- and GaAs-based VCSEL wafers are comparable and closely align with the designed wavelength of 940 nm. Increasing the substrate thickness causes 1-3 nm red shifts in both the stopband center wavelength and Fabry-Pérot dip toward longer wavelengths for both Ge and GaAs wafers. Larger substrate thicknesses are believed to cause different gas flows in and around the wafer pockets, which resulted in the red shifts.

Ge-based samples exhibit better control over QW layer thickness, with smaller deviations from the design compared to GaAs-based samples, indicating more precise and stable epitaxial growth. EDX analysis of elemental composition shows that the Ge330 sample has slightly smaller deviations from the stoichiometric value, suggesting a more consistent quantum well composition. Both Ge330 and GaAs625 exhibit similar strain distribution and comparable strain values, contributing to effective carrier confinement. Additionally, Ge-based VCSELs feature defect-free QW regions, with threading dislocation densities below $2.13 \times 10^6$ cm$^{-2}$.

Overall, Ge substrates offer advantages in better thermal contacts to the susceptors, better thickness control and composition precision, demonstrating more consistent epitaxial growth. These findings highlight Ge substrates as promising candidates for advanced VCSEL fabrication.

## Acknowledgment

The authors gratefully acknowledge LandMark Optoelectronics Corporation for the epitaxial growth services and technical support throughout the project. This work was supported by the Natural Sciences and Engineering Research Council of Canada (NSERC), the University of British Columbia (UBC), and the Southern University of Science and Technology (SUSTech) International PhD Fellowship. The Southern University of Science and Technology Core Research Facilities is acknowledged for providing the FIB and TEM imaging facility.